\documentclass[11pt,twoside]{article}

\usepackage{asp2004}
\usepackage{epsf}
\usepackage{psfig}
\usepackage{lscape}
\usepackage{graphicx}
\usepackage{natbib}
\def\deg{\hbox{$^\circ$}}
\def\newpar{\par\vspace{2mm} \noindent}

\def\ga{\mathrel{\hbox{\rlap{\hbox{\lower4pt\hbox{$\sim$}}}\hbox{$>$}}}}
\def\la{\mathrel{\hbox{\rlap{\hbox{\lower4pt\hbox{$\sim$}}}\hbox{$<$}}}}
\def\msunyr{$M$ \mbox{$_{\normalsize\odot}$}\rm yr^{-1}}
\def\kms{\,\rm km\,s$^{-1}$}
\def\msun{$M$\mbox{$_{\normalsize\odot}$}}

\def\rsun{$R$\mbox{$_{\normalsize\odot}$}}
\def\rstar{$R$\mbox{$_{\star}$}}

\def\eg{e.g.\,}

\bibpunct{(}{)}{;}{a}{}{,}

\begin{document}
\title{The clumpiness of LBV winds}   
\author{Ben Davies$^{1}$, Ren\'{e} D. Oudmaijer$^{1}$, Jorick S. Vink$^{2,3}$ }   
\affil{$^{1}$School of Physics \& Astronomy, University of Leeds, UK
  \\ $^{2}$Blackett Laboratory, Imperial College, London, UK \\
  $^{3}$Astrophysics Group, Lennard-Jones Laboratories, Keele
  University, UK}    

\begin{abstract} 
We present the first systematic spectro-polarimetric study of Luminous
Blue Variables (LBVs), and find that at least half those objects
studied display evidence for intrinsic polarization -- a signature of
significant inhomogeneity at the base of the wind. Furthermore,
multi-epoch observations reveal that the polarization is variable in
both strength and position angle. This evidence points away from a
simple axi-symmetric wind structure \`{a} la the B[e] supergiants, and
instead suggests a wind consisting of localised density enhancements,
or `clumps'. We show with an analytical model that, in order to
produce the observed variability, the clumps must be large, produced
at or below the photosphere, and ejected on timescales of days. More
details of LBV wind-clumping will be determined through further
analysis of the model and a polarimetric monitoring campaign.
\end{abstract}

\section{Introduction}

Luminous Blue Variables \citep[LBVs, ][ \& Nota these
proceedings]{H-D94} are very massive evolved stars that are found
right at the top of the HR diagram, roughly the same region as the
B[e] supergiants. Like the B[e]SGs, their spectra contain many
emission lines due to their strong winds. Unlike the B[e]SGs, they are
strongly photometrically variable, with amplitudes of 1-2 mags over
timescales of a few months to years. The amount of material ejected by
the LBVs is huge; their mass-loss rates are among the highest known
($10^{-5} - 10^{-4}\msunyr$), whilst they can throw off many solar
masses during giant eruptions \citep[\eg$\eta$ Car, ][]{Morris99}. The
LBV phase therefore represents a crucial stage in a massive star's
evolution.

\newpar

The products of the extreme mass-losing episodes of LBVs can be seen
in the form of their surrounding nebulae
\citep[\eg][]{Nota95}. However, it is unclear whether the bi-polar
nature of these nebulae is due to a pre-existing density contrast, or
if the wind itself is axi-symmetric. Such a wind may be linked to the
asphericity of supernova explosions and the beaming of gamma-ray
bursts; whilst wind asphericity is of major importance to stellar
evolution, as it can lead to large over-estimates of mass-loss rates
\citep[\eg ][]{Bouret05}.

\newpar 

The study of the inner-wind morphology of these stars is not
straight-forward. They are too far away to directly image the inner
wind, and spectroscopy alone gives no unambiguous geometry
information. The only tool capable of determining the present-day
mass-loss geometry is spectropolarimetry, and we present here a
comprehensive spectropolarimetric study of all LBVs in the Galaxy and
the Magellanic Clouds.

\subsection{Introduction to Spectropolarimetry}

\begin{figure}[h]
\centering
\includegraphics[width=8cm,angle=-90,clip]{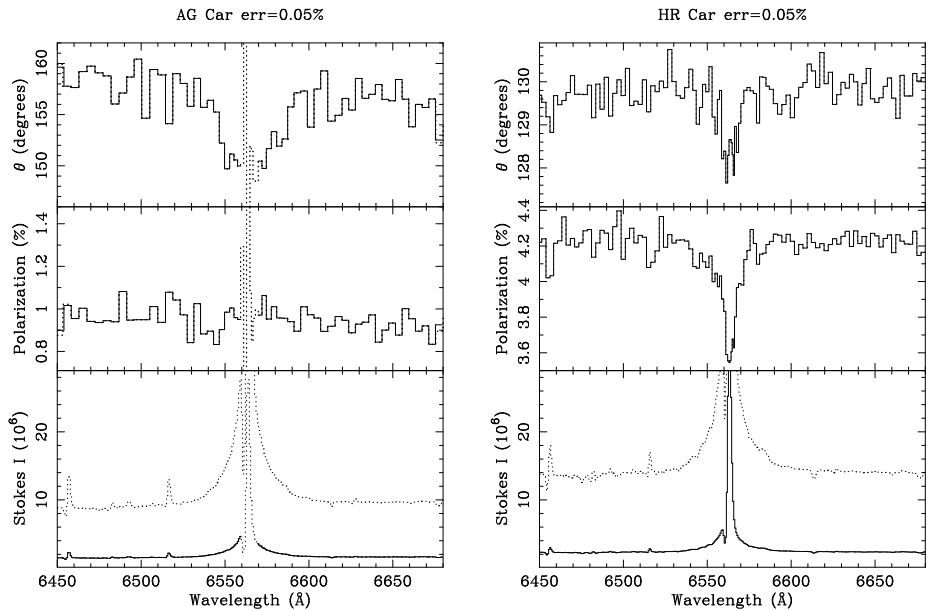}
\caption{Polarization spectra of two LBVs, AG Car ({\it left}), and HR
  Car ({\it right}). The bottom panels show the intensity spectrum in
  the region of H$\alpha$; the middle panel and upper panels show the
  degree of polarization and PA as a function of wavelength
  respectively. The changes in polarization across the stars' emission
  lines are indicative of aspherical structure low in the wind. Taken
  from \citet{Davies05}.}
\label{specpol}
\end{figure}

The technique is based on the following: the continuum emission of a
star is electron-scattered by the ionised material at the base of the
wind, while the line-emission, which forms in the wind over a much
larger volume, remains essentially unscattered. If the material at the
base of the wind is aspherical (\eg an equatorially-enhanced flow), the
continuum light will be left with a net polarization perpendicular to
the plane of scattering. The line-emission, which is unscattered,
remains unpolarized. The dilution of the polarized flux by the
line-emission can be seen as a drop in polarization across the
emission line (see Fig. \ref{specpol}). As the majority of the
polarization occurs within a couple of stellar radii
\citep{Cassinelli87}, by looking for changes in polarization across
strong emission lines such as H$\alpha$ we can look for inhomogeneous
structure at the base of the wind.


\section{Spectropolarimetric observations of LBVs}

Using the above-described technique, we have observed all known LBVs
in the Galaxy and Magellanic Clouds. We find that at least half of
those observed show changes in polarization across H$\alpha$, and
therefore signatures of asphericity at the base of the wind. We put
this 50\% detection rate as a lower limit, due to the difficulty in
achieving the target S/N of 1000 (0.1\% precision) for the fainter
stars. This compares with detection rates of 20\% and 25\% in similar
studies of Wolf-Rayet stars and O supergiants respectively, where
positive detections were very much the exception rather than the rule
\citep{Harries98,Harries02}.

\begin{figure}[h]
\centering
  \includegraphics[height=5cm,bb=0 445 300 702,clip]{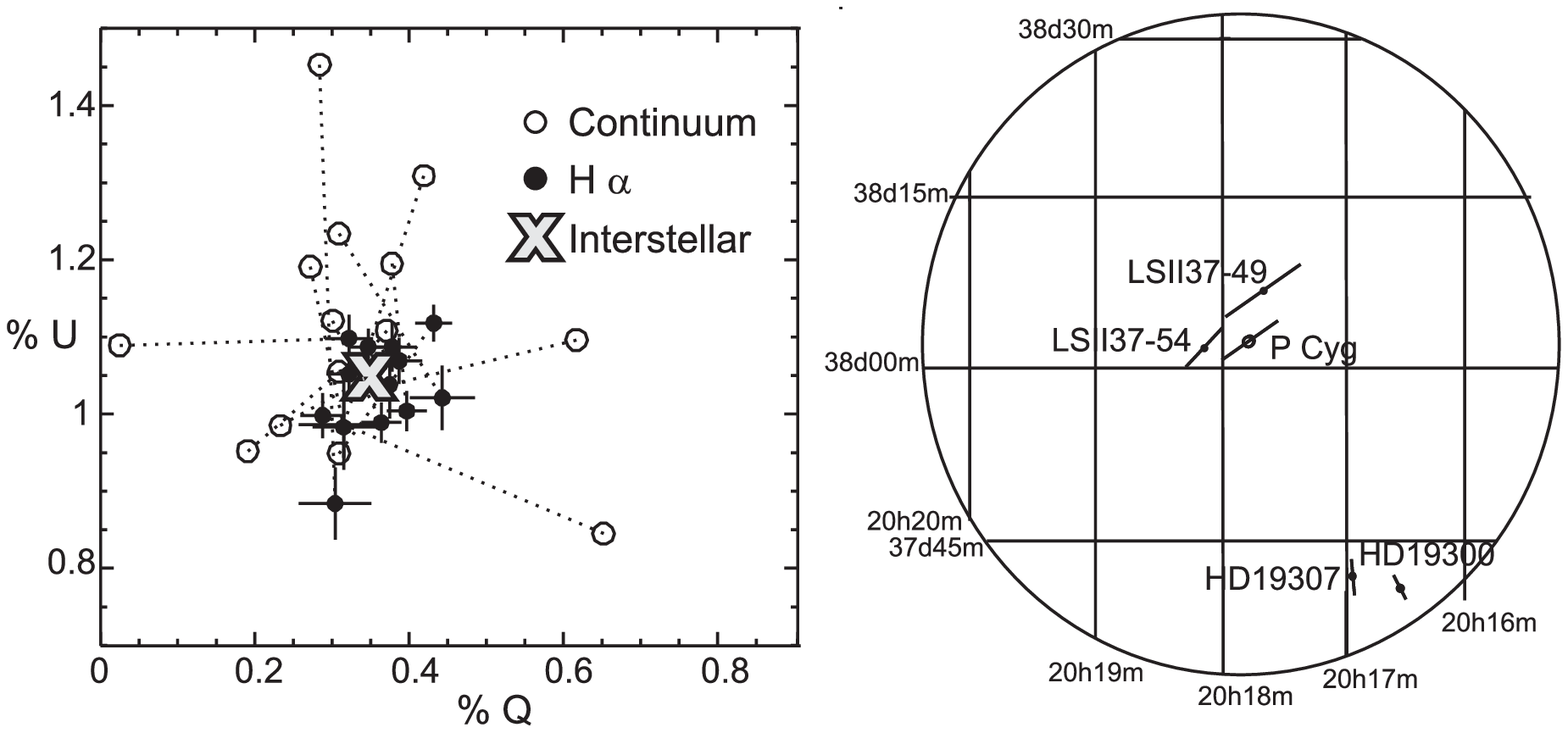}
  \includegraphics[height=5cm,bb=0 18 509 460,clip]{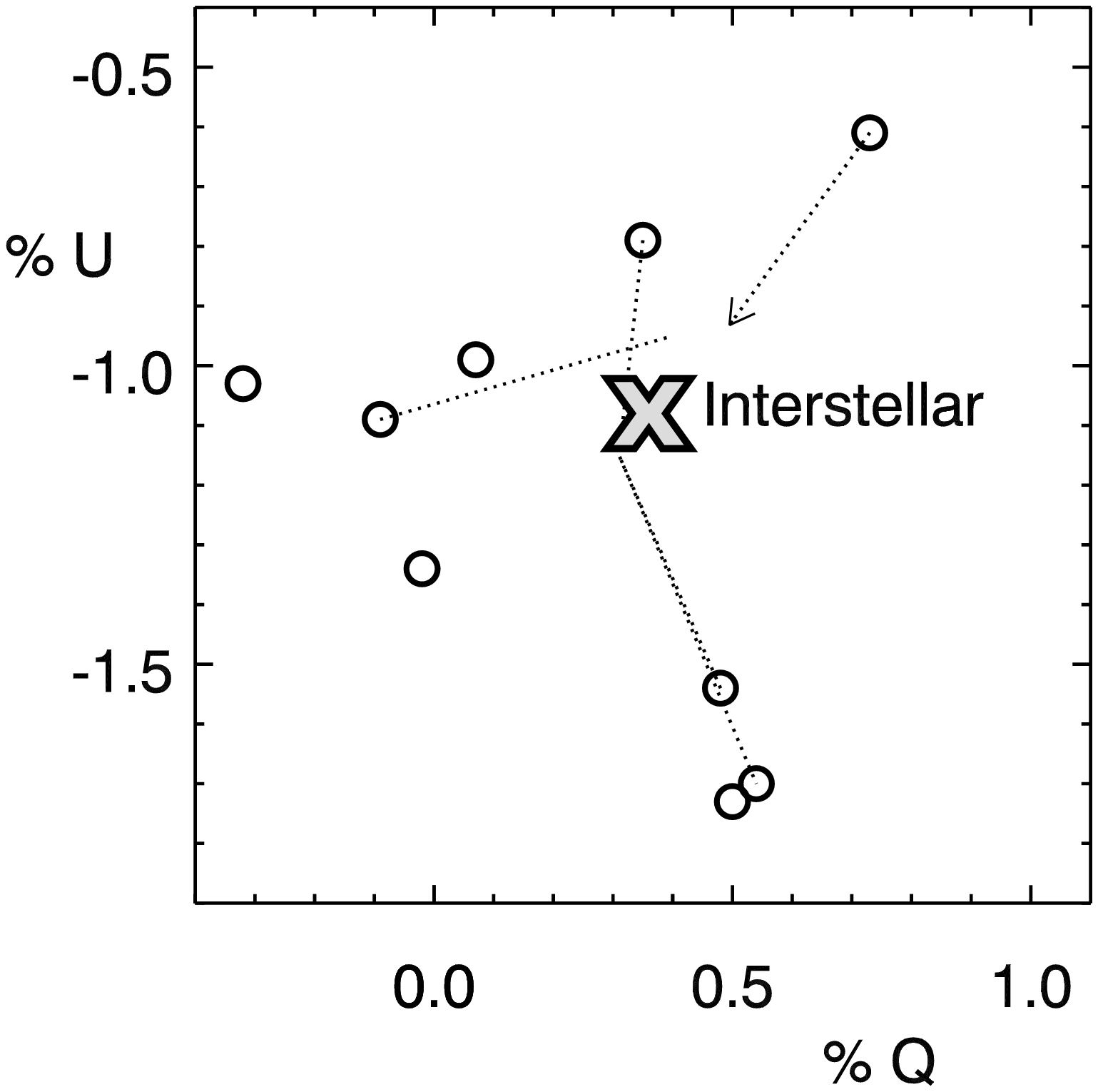}
\caption{{\it Left:} Polarization vector diagram of P Cyg. The
  magnitude of the vectors between the continuum measurements (open
  circles) and the zero-point polarization of P Cyg (marked
  `interstellar') represent the strength of the intrinsic
  polarization; while the angle of the vector with the $Q$ axis
  represents the polarization's position angle (PA). The polarization
  is variable in strength and PA on timescales of days. Taken from
  \citet{Nordsieck01}. \smallskip \newline
  {\it Right:} Same as the figure on the left
  but for AG Car. The zero-point polarization (marked `interstellar')
  has been determined from the H$\alpha$ depolarizations (dotted
  lines). Adapted from \citet{Davies05}.}
\label{epochs}
\end{figure}

In addition, we find that intrinsic polarization is more likely to be
found in those objects with strong line-emission, and those that have
been variable by $\ga$1 mag in the last $\sim$10 years
\citep[see][]{Davies05}. This may be linked to the fact that the
region occupied in the HR-diagram by the LBVs is close to certain
instabilities, namely the bi-stability jumps and the
Humphreys-Davidson limit. Strong variability in this regime means that
they can often appear to stray into these zones, possibly resulting in
erratic mass-loss behaviour, leading to strong line-emission and
wind-inhomogeneity.

\subsection{Temporal Variability}

\begin{figure}
\centering
\includegraphics[height=12cm]{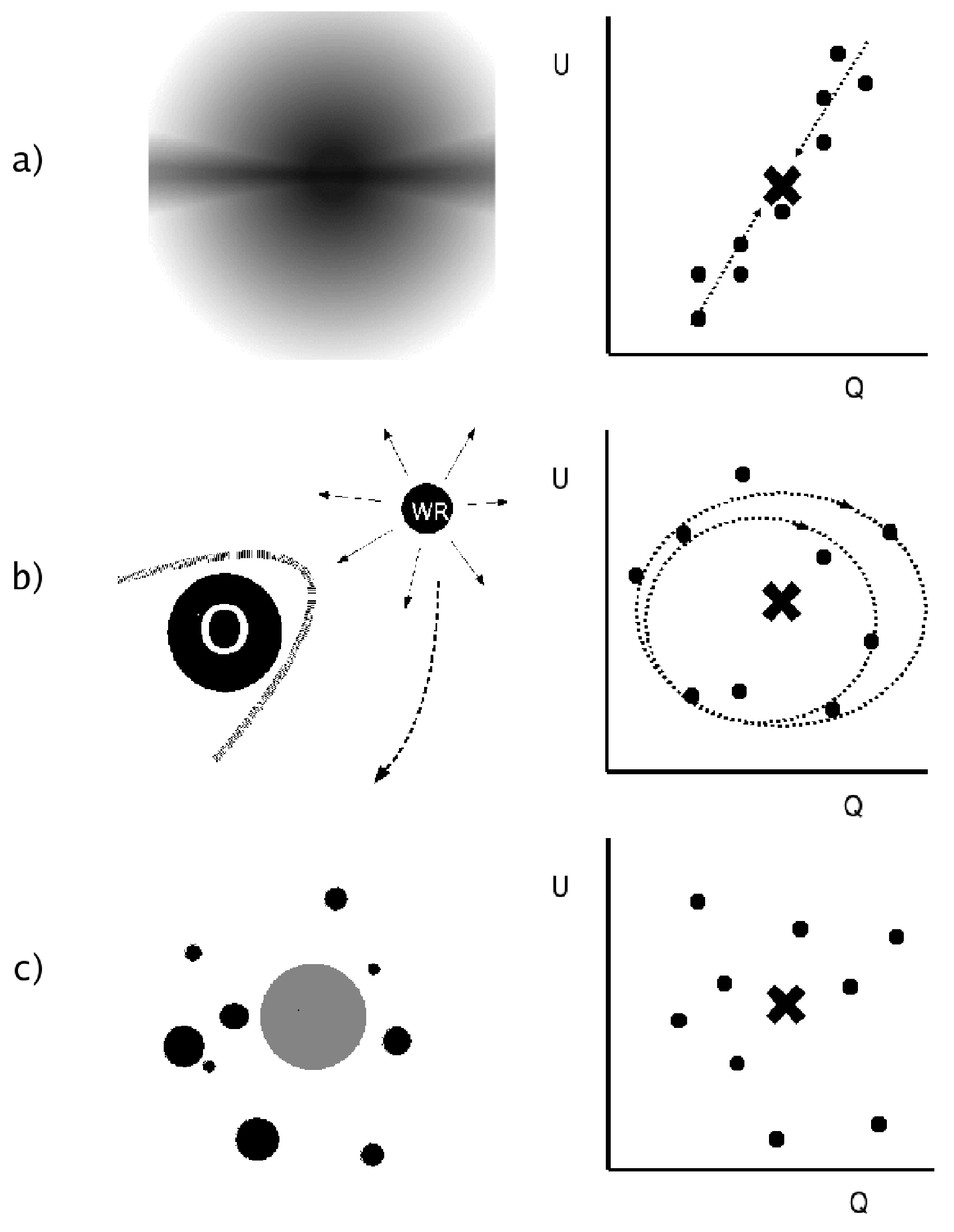}
\caption{Cartoons of three different scenarios capable of producing
  variable polarization ({\it left}), and the corresponding $Q-U$
  behaviour we would expect ({\it right}). Symbols are the same as in
  Fig. \ref{epochs}.  \newline
  {\bf (a):} an axi-symmetric wind. When the PA
  of polarization is perpendicular to the disk, we measure
  polarization along one vector in $Q-U$ space. If the PA were to
  rotate by 90\deg ~due to \eg an opacity increase at the equator or a
  flip from equatorial to bi-polar wind, we would measure polarization
  along a vector on the opposite quadrant of $Q-U$ space. \newline
  {\bf (b):} binarity, \eg a WR-O system. The O star's light is polarized
  due to scattering off the wind-wind collision zone. As the density
  enhancement region rotates with the binary system, the polarization
  describes a double-loop in $Q-U$ space.  \newline
  {\bf (c):} wind-clumping. Each clump in the wind scatters the starlight. If the
  clump distribution is not spherically symmetric, a net polarization
  results. As the clump distribution changes with time, so does the
  polarization.}
\label{cartoons}
\end{figure}

For four of our stars (AG Car, HR Car, P Cyg and R127) there exists
multi-epoch data in the literature and archives
\citep{S-L94,S-L93,Leitherer94,Clampin95,Taylor91,Nordsieck01}. For
all four stars, the intrinsic polarization they exhibit is variable in
both strength and PA (see Fig. \ref{epochs}). This is inconsistent
with a steady axi-symmetric wind, as observed in Be stars, where we
would expect the PA to remain constant. To explain the variable
polarization, we considered four possible scenarios, which are
summarized below (see Fig. \ref{cartoons}). For a full discussion on
each explanation and their merits, see \citet{Davies05}.

\begin{itemize}
\item An axi-symmetric wind with a variable optical depth, causing the
  polarized emission to originate from different latitudes as a
  function of time. In this scenario, polarization at any epoch would
  be aligned with one of two perpendicular directions (see
  Fig. \ref{cartoons}a). As this is not observed, this rules out this
  scenario as the sole cause of the variability.
\item A `flip-flopping' wind, where the wind's enhancement-plane
  switches between equatorial and bi-polar. This would again produce
  polarization in two perpendicular planes and can be ruled out. 
\item Binarity, similar to the mechanism producing the polarization
  observed in WR-O binary systems, see \eg \cite{StL87}. This is
  unlikely due to the physical properties required of the companion;
  it would need to have comparable luminosity and/or mass-loss rate to
  the LBV. It is unlikely such an object would have not already been
  detected through doppler-wobble, composite spectrum, or X-rays from
  the wind-wind collision zone.
\item A `clumpy' wind, where the polarization is caused by light
  scattering off localised density-enhancements low in the wind. As
  the clumps move away from the star in the wind and new clumps are
  constantly being formed, the distribution of clumps and hence the
  net polarization will appear very different from epoch to
  epoch. This would produce the stochastic polarization variations we
  observe in LBVs (cf.\,Fig. \ref{epochs}), and is the most likely
  explanation.
\end{itemize}

A clumpy inner wind is consistent with the clumpy nature of these
stars' inner nebulae \citep[\eg][]{Nota95,Chesneau00} and the
spectroscopic variability of their discrete absorption components
\citep[e.g.\,][]{Stahl01}. It is worth noting that wind-clumping is
also the favoured explanation of similar polarimetric behaviour seen
in WRs by \eg \citet[][]{Robert89}, who also find that the amplitude
of the variability {\it decreases} with {\it increasing} terminal wind
velocity. This is consistent with the LBVs, which have much slower
winds ($\sim$200\kms, compared with $\sim$1500\kms), with greater
polarimetric amplitude ($\sim$1\% compared with $\la$0.2\%).

\newpar

As an aside, the first few polarimetric observations of AG Car by
\cite{S-L94} suggested some form of axi-symmetry, leading the authors
to suggest one of the first two explanations. However, these scenarios
are not supported by subsequent observations
\citep{Leitherer94,Davies05}. This highlights the pit-falls in drawing
conclusions from single-epoch polarimetry data.

\section{Simulations of clumpy winds}

\begin{figure}
\centering
 \includegraphics[width=9.5cm,bb=12 250 460 730,clip]{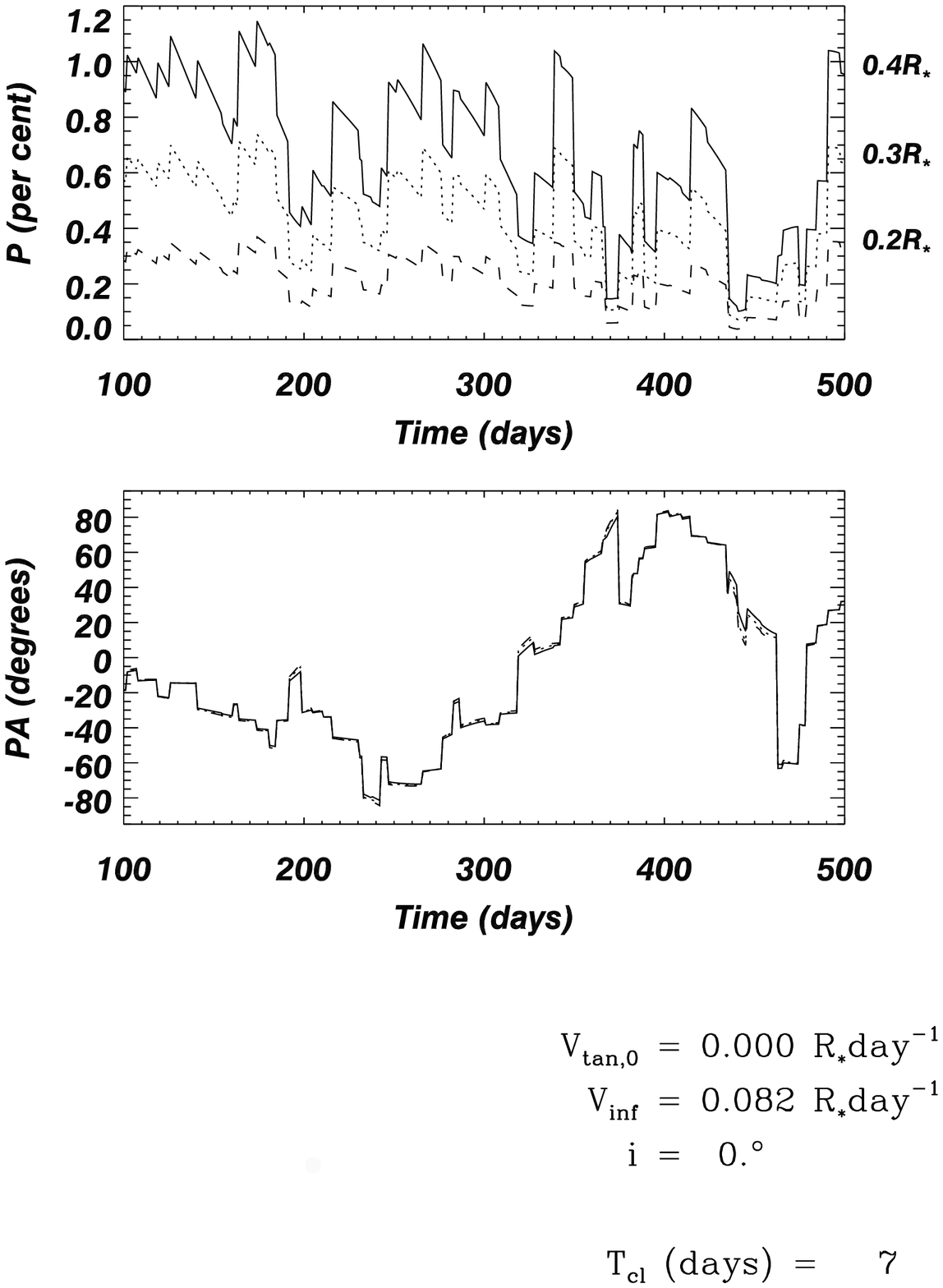}
\caption{Simulations of the time-dependent polarization produced by a
  random distribution of clumps in a spherical wind. The simulations
  use typical LBV parameters of $M = 30\msun$, $R = 300\rsun$, and
  $v_{\infty} = $200\kms; and the clumps have $\tau_{e} = 1$ and
  ejection timescale of 7 days. Runs using three different clump sizes
  of $0.4\rstar$, $0.3\rstar$, and $0.2\rstar$ are shown. From Davies
  (2005, in prep). }
\label{model}
\end{figure}

In order to begin to get a handle on typical clump parameters, we have
constructed an analytical model which simulates the polarimetric
variability due to wind clumping. Clumps are given a typical size,
optical depth, and ejection timescale. Once ejected radially from a
random position on the surface of the star, the polarization of one
clump is tracked as it accelarates with a $\beta=1$ velocity law. The
polarization due to all the clumps in the wind is calculated as a
function of time, as new clumps are ejected and move through the
wind. A typical run is shown in Fig.\,\ref{model}. Full analysis of
the models is pending (Davies et al. 2005 in prep), but initial
results are summarized below.

\newpar

It can be seen that, in order to produce the observed levels of
polarization and polarimetric variability, the clumps must be large
($\ga$ 0.2\rstar) and ejected every few days. Increasing the density
of the clumps (and hence the number of scatterers) does {\it not}
allow the clump size to be reduced, as Monte-Carlo studies have shown
that the polarization per clump plateaus at the $\tau_{e} \ga 1$ level
\citep{R-M00}. Also, the polarization per clump falls off dramatically
with distance. This means that {\it (a)} the clumps must be produced
at or below the photosphere, not condense out of the wind at larger
distances due to radiative instabilities \citep[\eg][]{R-O02}; and
{\it (b)} there must be a constant supply of clumps.

\newpar

For a given mass-loss rate and density contrast between clump and
ambient wind \citep[$\sim$ 20, ][]{Nordsieck01}, the polarization as a
function of clump size and ejection timescale can be
simulated. Through polarimetric monitoring of the LBVs we will
constrain the clump ejection timescale through the signature of jumps
in polarization that ejection events cause (Fig. \ref{model}). From
this, we can ultimately determine the {\it volume filling-factor} of
the clumps, the parameter to which wind models are so sensitive
\citep{Bouret05}.

\section{Summary}

In the first comprehensive spectropolarimetric study of LBVs, we find
that at least 50\% show evidence of intrinsic polarization -- over
double the detection rate of WRs and O supergiants. Moreover, the
polarization is stochastically variable, suggesting that the
polarization is not due to axi-symmetric wind structure but instead
caused by significant wind-clumping. Preliminary modelling of this
shows that the clumps must be very large, and produced at/below the
photosphere -- clumps condensing out of the wind at large distances
cannot produce the observed variability. Polarimetric monitoring will
allow us to constrain further typical clump parameters, and ultimately
determine the wind clump filling-factor, enabling more accurate
mass-loss rate estimates.





\end{document}